\documentclass[12pt]{article}
\newcommand{\be}[3]{\begin{equation}  \label{#1#2#3}}

 \input psfig
\newcommand{\ee}{ \end{equation}}
\newcommand{\ba}{\begin{array}}
\newcommand{\ea}{\end{array}}

\begin{document}

\thispagestyle{empty}
\rightline{UPR-857-T, NSF-ITP-99-107,CALT-68-2240}
\rightline{\it revised version}
\rightline{hep-th/9909058}

\vspace{1truecm}

\centerline{\bf \Large 
Supersymmetric Domain Wall World from}

\vspace{.5truecm}

\centerline{
\bf \Large D=5  Simple Gauged Supergravity
}
\vspace{1.2truecm}
\centerline{\bf Klaus Behrndt$^a$\footnote{e-mail: 
 behrndt@theory.caltech.edu}
{\rm and} Mirjam Cveti{\v c}$^b$\footnote{e-mail: cvetic@cvetic.hep.upenn.edu}}
\vspace{.5truecm}
\centerline{$^a$ \em California Institute of Technology}
\centerline{\em Pasadena, CA 91125}

\vspace{.3truecm}

\centerline{$^b$ \em Department of Physics and Astronomy}
\centerline{\em University of Pennsylvania, Philadelphia, PA 19104-6396}

\centerline{and}

\centerline{\em Institute for Theoretical Physics}
\centerline{\em University of California,  Santa Barbara, CA 93106}

\vspace{1truecm}

%%%%%%%%%%%%%%%%%%%%%%%%%%%%%%%%%%%%%%%%%%%%%%%%%%%%%%%%

\begin{abstract}
We address a supersymmetric embedding of domain walls with asymptotically
anti-deSitter (AdS) space-times in five-dimensional simple, N=2 U(1)
gauged supergravity theory constructed by Gunaydin, Townsend and Sierra.
These conformally flat solutions interpolate between supersymmetric AdS
vacua, satisfy the Killing spinor (first order) differential equations,
and the four-dimensional world on the domain wall is a flat world with
N=1 supersymmetry. Regular solutions in this class have the energy density
related to the cosmological constants of the supersymmetric AdS vacua. An
analysis of such solutions is given for the example of one (real,
neutral) vector supermultiplet with the most general form of the
prepotential.  There are at most two supersymmetric AdS vacua that are in
general separated by a singularity in the potential. Nevertheless the
supersymmetric domain wall solution exists with the scalar field
interpolating continuously across the singular region.
% As the gauge coupling becomes very
% large (compared to five-dimensional Planck constant), these domain walls
% become infinitely thin, and a special case of a $Z_2$ symmetric domain
% wall is a supersymmetric realization of the static domain wall solution
% considered by Randall and Sundrum.
\end{abstract}

%%%%%%%%%%%%%%%%%%%%%%%%%%%%%%%%%%%%%%%%%%%%%%%%%%%%%%%%%%%%%%%%%%%%%

\newpage

%%%%%%%%%%%%%%%%%%%%%%%%%%%%%%%%%%%%%%%%%%%%%%%%%%%%%%%%%%%

\section{Introduction}
The past few months have witnessed exciting progress in the study of
domain walls  in D=5 gravity theories. Such configurations are
interesting from two, on a surface orthogonal perspectives: (i) in the
context of AdS/CFT correspondence such conformally flat configurations
provide new insights in the study of RGE flows
\cite{010,190,180,220,020,150,160,030,230} and (ii) in the context of
phenomenological implications, such configurations provide a framework
\cite{050,070,060,240,250,260,211,210,212} to address the physics
implications of large dimensions for the four-dimensional world on the
domain wall.

Within the first approach a number of conformally flat solutions were
constructed and in particular the ones interpolating between
supersymmetric anti-deSitter (AdS) vacua of N=8 D=5 gauged theory
provide examples of static domain walls in D=5 with implications for
the renormalization group flow and spectra in strongly coupled
four-dimensional super Yang-Mills theories. One such example
\cite{190,020} involves two scalar fields and thus was solved
numerically and another most recent example with one scalar field can
be solved explicitly \cite{261}.

Within the second approach infinitely thin, static, $Z_2$-symmetric
domain wall solutions were constructed \cite{050,070} for pure AdS
gravity theory. (Generalizations that incorporate effects of additional
compactified dimensions were given in \cite{060,250,260}.) These
solutions have to satisfy a specific relation between the domain wall
tension $\sigma$ and the cosmological constant $\Lambda$ of the AdS
vacua, thus implying a fine-tuning.

The purpose of this letter is few-fold. We shall address a supersymmetric
embedding of domain walls with asymptotically AdS space-times in
five-dimensions, in the simplest supergravity theory, i.e. the
supergravity theory with least supersymmetry that allows for the explicit
constructions of supersymmetric domain wall configurations. In order to
demonstrate the existence of such domain walls, a supergravity theory
necessarily has to have a potential for (gauge neutral) scalar fields,
and the only known such examples are gauged supergravity theories and the
matter fields responsible for the formation of the domain wall belongs to
vector-supermultiplets. We thus choose to work within a framework of a
five-dimensional N=2, U(1) gauged supergravity formulated by Gunaydin,
Townsend and Sierra \cite{080}~\footnote{Such a theory may have its
origin as a compactification of Type IIB superstring theory on a specific
Einstein-Sasaki-5-manifolds or as a compactification of M-theory on
Calabi-Yau with non-trivial fluxes turned on. But the details are unknown
so far and remain to be worked out. }.

We derive the Killing spinor equations for domain wall solutions. For
regular solutions in this class we also derive that the energy density
$\sigma$ of the wall is related in a specific way to the cosmological
constants of the supersymmetric vacua on each side of the
wall. Namely, the relationship between the domain wall tension and the
cosmological constant for regular supersymmetric domain walls is a
consequence of the BPS nature of the solution, and not an artifact of
fine-tuning~\footnote{These properties are very much parallel to those
of supersymmetric domain walls of four-dimensional N=1 supergravity
theory found in \cite{120} and reviewed in \cite{130}.}.  These
configurations have four unbroken supercharges, or in other words
break $1\over 2$ of N=2, D=5 supersymmetry, and thus the
four-dimensional world on the domain wall has N=1 supersymmetry. A
special example of infinitely thin supersymmetric domain walls
(corresponding to the case of very large gauge coupling) with $Z_2$
symmetry would provide a concrete supersymmetric realization of the
static domain wall solution found by Randall and Sundrum \cite{050}.
 
For the sake of concreteness we analyse the case with one physical
(gauge neutral) vector superfield which allows for an explicit
analysis of the possible domain wall configurations. For this simple
model, kink solutions have been discussed some time ago \cite{081};
the domain walls presented in this paper provide a concrete and
explicit realisation of supersymmetric kink solutions. The upshot of
the analysis is that this framework does provide examples of static
domain wall solutions that satisfy the Killing spinor equations and
are thus supersymmetric.  The general one scalar case, however, has a
potential that has at most two supersymmetric AdS vacua which are
always separated by at least one (out of three) singularity points
where the potential diverges. But nevertheless the scalar for the
supersymmetric domain wall solutions interpolates across the singular
point.

%%%%%%%%%%%%%%%%%%%%%%%%%%%%%%%%%%%%%%%%%%%%%%%%%%%%%%%%%

\section{D=5 N=2 U(1) Gauged Supergravity}

%%%%%%%%%%%%%%%%%%%%%%%%%%%%%%%%%%%%%%%%%%%%%%%%%%%%%%%

Supergravity in D=5 is very restrictive with respect to allowed
potentials. The only allowed potentials come from gauging of
isometries and especially interesting are potentials that have no
``run-away'' behavior (scalars become asymptotically constant) with
non-trivial isolated extrema.  This type of potential allows for the
existence of domain walls with extrema corresponding to the AdS vacua
on each side of the wall.  The minimal gauged supergravity (N=2 gauged
supergravity with $U(1)$ gauged $R$-symmetry), constructed in
\cite{080,081}, provides such a set-up. In this case one can
consistently decouple the hyper-multiplets and the Lagrangian contains
only the supergravity multiplet and the vector supermultiplets.
(There are also domain wall solutions, that couple to non-trivial
hypermultiplets \cite{211}, but they do not have asymptotic anti-de
Sitter spaces.)  In this case the bosonic Lagrangian reads:
\begin{equation}
S_5 = \int \Big[{\frac{1}{2}} R + g^2 V - {\frac{1}{4}} G_{IJ}
F_{\mu\nu} {}^I F^{\mu\nu J}-{\frac{1}{2}} g_{AB} \partial_{\mu}
\phi^A \partial^\mu \phi^B \Big] + {1 \over 48} \int C_{IJK} F^I
\wedge F^J \wedge A^K \label{Lag} 
\ee 
We chose the convention where
the five-dimensional Newton's constant is $\kappa=1$ and $g$ is the
gauge coupling.  We work in the $(-,+,+,+,+)$ convention.  The
physical scalars $\phi^A$, which are real and neutral, correspond to
the scalar components of the vector super-multiplets and define
coordinates of the manifold defined by \cite{080}
\begin{equation}
F = {1 \over 6} C_{IJK} X^I X^J X^K = 1   \ ,  \label{const}
\ee
with $C_{IJK}$ real, and the $X^I$ are the auxiliary real scalar fields. The
  metric(s)  of the scalar manifold $g_{AB}$ (for physical scalars $\phi^A$)
   and $G_{IJ}$  (for auxiliary scalars $X^I$) are defined by
\begin{equation}
G_{IJ} = - {1 \over 2} \Big(\partial_{I} \partial_J \, \log {F}
\Big)_{F = 1} 
\qquad , \qquad  
g_{AB} = \Big(\partial_A X^I \partial_B X^J G_{IJ} \Big)_{F=1} 
\ee
where $\partial_A \equiv {\partial \over \partial \phi^A}$.
$\partial_I \equiv {\partial \over \partial X^I}$.  The  auxially scalars $X^I$
are accompanied by   gauge field strengths 
$F^I_{\mu\nu}$ entering  the Lagrangian (\ref{Lag}).

The gauging of a $U(1)$ subgroup of the $R$-symmetry introduces a
potential for the scalars \footnote{Note the parallels with the potential
in D=4 N=1 supergravity where: 
$V=e^K \Big( g^{A\bar B} D_A W \overline{D_B W} - 3 | W |^2 \Big)
$ where $W$ and $K$ are the
superpotential and K\" ahler potential for the chiral superfields.}
\begin{equation}     \label{potential}
\ba{rcl}                     
V &=& 6\, h_I h_J \Big( X^I X^J - {3 \over 4} 
g^{AB} \partial_A X^I \partial_B X^J \Big) \\
&=& 6\, \Big( W^2 - {3 \over 4} g^{AB} \partial_A W \partial_B W \Big) \ ,
\ea
\ee
where $h_I$ are real constants, specifiying the Fayet-Iliopoulos(FI) terms,
 and the superpotential $W$ is defined as
\begin{equation}
W = h_I X^I \ . \label{super}
\ee
Notice, $W$ is subject to the constraint (\ref{const}) which makes it
non-linear in the physical scalars $\phi^A$.

%%%%%%%%%%%%%%%%%%%%%%%%%%%%%%%%%%%%%%%%%%%%%%

\subsection*{Supersymmetry Transformations and BPS-Saturated Domain Walls}

%%%%%%%%%%%%%%%%%%%%%%%%%%%%%%%%%%%%%%%%%%%%
We are searching for supersymmetric (BPS-saturated) domain wall solutions:
those are solutions that preserve part of the supersymmetry, and thus
satisfy the Killing spinor equations, which are first order
differential equations what ensure that the supersymmetry
transformations in this domain wall background are preserved.

We choose these domain wall solutions  to be neutral, and thus they
are  supported only by (gauge neutral) scalars with the gauge fields  turned
off. Thus, the supersymmetry transformations  for these backgounds read
\cite{080}:
\begin{equation}
\ba{l}
\delta \lambda_A = \Big( - {i \over 2} g_{AB} \Gamma^{\mu}
\partial_{\mu} \Phi^B + i\, {3 \over 2} g  \partial_A W \Big)
\epsilon \ ,\\
\delta \psi_{\mu} = \Big( \partial_{\mu} + 
{1 \over 4} \omega_{\mu}^{ab} \Gamma_{ab}
+ {1 \over 2} g \, \Gamma_{\mu} W \Big) \epsilon \ .
\ea
\ee
The vacuum is given by the asymptotic space, where the scalars are
constant and thus supersymmetry requires $\partial_A W=0$. The form
of the potential $V$ (\ref{potential}) implies that supersymmetric
vacua are always extrema of the potential.

The domain wall Ansatz for the metric is of the form:
\begin{equation}
ds^2= A(z)\Big[-dt^2 + dx_1^2+dx_2^2+dx_3^2\Big] + dz^2 \ , \label{metri}
\end{equation}
and the scalars have the form $\phi^A=\phi^A(z)$, where $z=\{-\infty
,+\infty\}$ is a direction transverse to the wall.

Then the Killing spinor equations $\delta \psi_{\mu} = 0$ and 
$\delta  \lambda_A = 0$ are solved by~\footnote{We would like to thank S. 
Gubser for pointing out the sign error in eq. (\ref{killm}) in the original
version of the manuscript. As  a consequence the  explicit
solution for the metric coefficient $A(z)$,  discussed in
Section 3 of the original version, becomes $A(z)^{-1}$ of the revised version.} 
: \begin{equation} \partial_z \log A = 2 g
W \ , \label{killm} \end{equation} and
\begin{equation}
\partial_{z} \phi^A = -3 g g^{AB} \partial_B W \ , \label{killp}
\end{equation}
where the four component spinor satisfies the constraint: 
$\Gamma_z \epsilon = - \epsilon$. (Killing spinor equations for 
domain walls of D=5, N=8 supergravity can be cast in a similar form
\cite{020}.)
Note that as long as the domain of physical fields contain two
isolated supersymmetric vacua, 
%(subject to the constraint that
%$g^{AB}\partial_B\partial_C W$
this set of solutions specify the BPS
domain wall. The physical domain of such solutions requires
that the scalar metric $g_{AB}$ remains positive definite.
(In D=4 N=1 supergravity,
the Killing spinor equations are similar \cite{120}:
 $\partial_z {\rm log} A\sim
e^{K\over 2} W$\ , $\partial_z\phi^A \sim  
e^{K\over 2}g^{A{\bar B}}{\overline {D_BW}}$.)

The domain wall tension can be determined by applying Nester's
procedure  which relates the wall tension $\sigma$ to the
central charge of the supersymmetry algebra; the central charge is
determined by the values of the superpotential at each asymptotically
supersymmetric vacuum. (For D=4  N=1 domain wall solutions, see Appendix A of
\cite{120}.) More concretely, one considers
the integral over the spatial boundary 
\begin{equation}
\int_{\partial \Sigma} N^{\mu\nu} d\Sigma_{\mu\nu} 
= \int_{\Sigma} \nabla_{\mu} N^{\mu\nu} d \Sigma_{\nu}
= \int_{\Sigma} \nabla_{\mu} N^{\mu 0} d \Sigma_{0}\  . \label{nester}
\ee
 $\Sigma_{\mu\nu}$ is a space-like 
hypersurface and thus $d \Sigma_{0} \sim dz d\vec x$. The  
Nester tensor reads $N^{\mu\nu} = \bar \epsilon
\Gamma^{\mu\nu\lambda}\delta \psi_{\lambda}$ where $\delta \psi_{\lambda}$
is the  gravitino variation. In (\ref{nester}) 
 we used the Stokes theorem, and thus assumed that the Nester tensor is
non-singular.  In order to determine the energy density, 
we can factor out the integral over
the domain wall coordinates ($d\vec x$) and the integration  over the 
transverse
direction ($z$) yields  in (\ref{nester}) the  contributions far away from the
wall.  Inserting the gravitino variation in (\ref{nester}), one
obtains two contributions, the first one represents the domain wall
tension $\sigma$ (energy density) and the second one corresponds to the
central charge ${\cal C}$. The latter one  is a topological term that
 corresponds to
the difference of the  boundary values of
the superpotential. For the  
supersymmetric configuration the gravitino variation is zero, and thus
 (\ref{nester}) is zero which implies:
\begin{equation}
\sigma \, 
%\equiv \Gamma^0 (-{1 \over 12} \sqrt{A}\, ')_{+\infty}^{-\infty}  \, 
 = {\cal C}\equiv 
- {1 \over 2} (\bar\epsilon \sqrt{A} \Gamma^0 \epsilon) \, g W
\Big|_{-\infty}^{+\infty} \ , \label{bog}
\ee
where we used the projector $\Gamma_z \epsilon = - \epsilon$. 
%In the  Killing spinor equations   and in (\ref{bog}) we chose 
%$W(\phi^A|_{z=+\infty}) \ge  W(\phi^A|_{z=-\infty})$. 
%(The reversal of this
%inequality implies the change of signs in the Killing spinor equations, 
%(\ref{bog}) and  $\Gamma_z \epsilon=\epsilon$.) 
Again, in the derivation of
(\ref{bog}) it was
assumed   that the Stokes theorem can be applied and thus the
 $\partial_z \ W$ is  non-singular inside the wall.
In the one-scalar example  this is not the case (see later).

Normalizing the Killing spinor as $(\bar \epsilon \Gamma^0 \sqrt{A}
\epsilon) = 1$, yields the result:
\begin{equation}
\sigma_{BPS}= - {g \over 2}  \Big(
W_{+\infty} - 
W_{-\infty}
\Big)
= -
{1\over 2 \sqrt{6}} 
\Big({\rm sign}[ W_{\infty}]\sqrt{- \Lambda_{+\infty}}-{\rm sign}[W_{-\infty}]
 \sqrt{- \Lambda_{-\infty}}
\Big) 
\ , \label{bogbo}
\end{equation}
where $W_{\pm\infty}\equiv W(\phi^A|_{z=\pm\infty})$. In the second
part of (\ref{bogbo}) we have used the relationship between the
cosmological constant $\Lambda$ and the value of the superpotential
$W$ for supersymmetric vacua. 
%Note that $\pm$ is determined by ${\rm sign} [ W_{+\infty}W_{-\infty}]$.
  Thus, the domain wall tension is
specified by the values of the cosmological constants of the asymptotic
AdS vacua. 

According to the asymptotic bahavior of the Killing spinor equation
(\ref{killm}) there are the following types of BPS-saturated domain walls
(very much parallel to the analysis of the types of BPS-saturated domain walls
in D=4 \cite{140}, their global space-time structure \cite{280} and their
relationship to non-supersymmetric configurations \cite{200}):
\begin{itemize}
\item
Type I domain walls interpolate supersymmetric Minkowski space-time
($\Lambda_{-\infty}=0$)
%, $W(\phi^A|_{-\infty})=0$) 
and the AdS space time ($\Lambda_{+\infty} \equiv \Lambda \ne 0$).
 On the asymptotic AdS side the metric coefficient takes
the form $ A(z)= e^{-\sqrt{-{{2 \over 3} \Lambda} }\, z}$; on
the AdS side of the wall $z\to \infty$ limit corresponds to the AdS horizon.
The geodesic extension of these space-times could either be pure AdS or
new regions that involve an infinite tiling with the ``mirror'' domain
walls.  Regular solutions have $\sigma_{BPS}= {1 \over 2 \sqrt{6}} 
\sqrt{-\Lambda}$. These
walls saturate an analog of the Coleman-De Luccia bound \cite{100} in five
dimensions. 
\item
Type II domain walls interpolate between supersymmetric AdS vacua
where ${\rm sign}[W_{-\infty}]=-{\rm sign}[W_{+\infty}]$  and the metric 
behaves as $A(z)=e^{- \sqrt{-{{2 \over 3} \Lambda_{\pm\infty}}}\, |z|}$
for  $z\to \pm\infty$, i.e.\  one approaches the AdS horizons 
and thus the geodesic extensions  could be either pure AdS or
new regions that comprise of an infinite tiling with the mirror domain
walls. For regular solutions
$\sigma_{BPS}=
{1\over 2 \sqrt{6}} \Big(\sqrt{- \Lambda_{+\infty}} + \sqrt{-
\Lambda_{-\infty}} \Big)$. These domain walls can be viewed as ``stable'';
non-supersymmetric generalizations are expanding bubbles on either side
of the AdS vacua \cite{270}.  A special case of a $Z_2$ symmetric solution
($W_{+\infty}=-W_{-\infty}$) satisfies the constraint:
$\sigma_{BPS}={1 \over 2 \sqrt{6}} \sqrt{\Lambda}$ which is
a relationship found in \cite{050}.
\item
Type III domain walls are those between two supersymmetric AdS vacua
 where ${\rm sign}[W_{-\infty}]=+{\rm sign}[W_{+\infty}]$. 
 The metric coefficient grows exponentially on one
 side of the wall: $ A(z)=e^{+{\rm
 min}[ \sqrt{-{{2 \over 3} \Lambda_{\pm\infty}}} ]\, |z|}$, and thus  on this
 side, $|z|\to \infty$  limit corresponds to the boundary of the AdS space!
 For regular solutions
 $\sigma_{BPS}=
 {1\over 2 \sqrt{6}} \Big(
\sqrt{- \Lambda_{+\infty}} - \sqrt{-
 \Lambda_{-\infty}} \Big)$; those are the ``unstable'' 
domain wall solutions whose
 non-supersymmetric generalizations corresponds to false vacuum decay
 bubbles, only \cite{270}.
\item
Type IV domain walls   correspond to a class of solutions where  ${\rm
sign}[W_{-\infty}]=-{\rm sign}[W_{+\infty}]$ (just like Type II walls), but
now the metric  behaves asymptotically as $A(z)=e^{\sqrt{-{{2 \over 3}
\Lambda_{\pm\infty}}}\, |z|}$, i.e. for  $|z|\to \infty$ one
approaches the boundary of the AdS space instead the horizon. 
They have negative surface density. 
%Their surface density is negative. 
A special limit in
this class (Type V) would correspond to the case where one, say,
$\Lambda_{-\infty}=0$.
\end{itemize}

%Similarly, there is a further type where the AdS side of type I does not
%approach the Killing horizon, but the AdS boundary.

%%%%%%%%%%%%%%%%%%%%%%%%%%%%%%%%%%%%%%%%%%%%%%%%%%%%%%%%%%%%%

\section{BPS Domain Walls with One Vector Supermultiplet}

%%%%%%%%%%%%%%%%%%%%%%%%%%%%%%%%%%%%%%%%%%%%%%%%%%%%%%%%%%%%%%

For the sake of being explicit we will address the
case of a single vector multiplet. Defining the physical scalar
as $\phi = X^1 / X^0$ the constraint
(\ref{const}) takes the form:
\begin{equation}
F=(X^0)^3 \, (A+B\phi+C\phi^2+D\phi^3)=1 \ ,  \label{const1}
\end{equation}
and  the superpotential (\ref{super})becomes:
\begin{equation}
W=X^0(h_0+h_1\phi)\ .\label{super1}
\end{equation}
where $X^0$ is the auxiliary field eliminated by eq.\ (\ref{const1}).
The metric $g_{\phi \phi}$, and the derivative of the potential
$\partial_\phi W$ take the form:
\begin{equation}
g_{\phi \phi}= {1 \over 3} {(C^2 - 3 BD) \phi^2 + (BC -9 AD) 
\phi + (B^2 -3AC) \over 
(A + B \phi  + C \phi^2 + D \phi^3 )^2 }\ , 
\end{equation}
\begin{equation}
\partial_\phi W= {({1 \over 3} h_1 C - h_0 D) \phi^2 + 
{2 \over 3} (h_1 B - h_0 C) \phi + h_1 A - {1 \over 3} h_0 B \over
(A + B \phi + C \phi^2 + D \phi^3)^{4/3}}\ ,
\end{equation}
and the potential reads:
\begin{equation}
V=6 \Big[W^2-{3\over{4g_{\phi \phi}}} (\partial_\phi W)^2 \Big]\ .
\end{equation}

One can make the following general observations about the nature of
supersymmetric vacua. (i) The superpotential (\ref{super1}) allows for
at most two extrema, where $\partial_{\phi} W = 0$. (ii) Expanding $W$
around a given extremum yields: $\partial_{\phi}^2 W|_{extr} =
\textstyle{2\over 3}\, g_{\phi \phi}\, W|_{extr}$ (see e.g.,\
\cite{150}).  This relationship implies that for physical solutions
with $g_{\phi\phi} >0$ the two extrema of $W$ cannot be connected,
there is at least one pole between them. Thus, the supersymmetric
domain wall solution necessarily involves a ``jump'' over a region
where the superpotential (as well as the scalar metric and the
potential) has a pole.  Note, these lines of arguments hold for the
one-scalar case, only. If $W$ depends on more than one scalar, it may
allow for two minima, which can be smoothly connected.

In order to  discuss the solution in more detail we choose, without loss of
generality,
 the following parameterization: 
\begin{equation}
g=1, \ \quad 
A=0\ , \quad B=D=h_0= 1\ , \quad C= \sqrt{3} \chi\ , \quad
h_1 = \sqrt{3} \xi\ .
\ee
(One can show that $g=D=h_0=1$ corresponds to an overall rescaling of
 the potential, $A=0$  can be obtained by 
shifts   $\phi\to \phi-\phi_0$  and $h_0\to h_0+h_1\phi_0$,  and  
$B=1$ corresponds to a rescaling of  $\phi$.) 
In this case the metric, superpotential and its derivate can be
written in the following form:
\begin{equation}
g_{\phi \phi}={{3(\chi^2-1)\phi^2+\sqrt{3}\chi\phi+1}
\over{3\phi^2(1+\sqrt{3}\chi\phi+\phi^2)^2}} \ ,
\label{g11}\end{equation}
\begin{equation}
W={{1+\sqrt{3}\xi\phi}\over{[\phi(1+\sqrt{3}\chi\phi+\phi^2)]^{1\over
3}}}\ ,
\label{w1}\end{equation}
\begin{equation}
\partial_\phi W=
{{3(\chi\xi-1)\phi^2-2\sqrt{3}(\chi-\xi)\phi - 1  }\over
{3\, [\phi(1+\sqrt{3}\chi\phi+\phi^2)]^{4\over 3}}}\ .
\label{delw1}\end{equation}
The corresponding Killing spinor equations for the metric coefficient
$A(z)$ is given in (\ref{killm}) and that for the scalar field
(\ref{killp}) takes the form:
\begin{equation}
g_{\phi \phi} \partial_z \phi= - 3 \, \partial_\phi W \ . \label{killp1}
\end{equation}

Let us mention that in a proper coordinate system, these first
order differential equations are solved by an algebraic constraint
\cite{150}, which says that the normal vector on the
scalar manifold, defined by (\ref{const1}), has to behave monotonically and
becomes parallel to $h_I$ in the asymptotic AdS vacuum.

\subsection*{Features of the Solutions and an Example}

The first useful observation is that in the region where the metric
$g_{\phi \phi}$ has real poles, $g_{\phi \phi}$ has no real zeroes.
Namely, the poles and zeros are at the following values of $\phi$:
\begin{equation}
{\rm poles\ \ of\  \ 
g_{\phi \phi}}: \ \ \{\textstyle{1\over 2}(-\sqrt{3}\chi \pm 
\sqrt{3 \chi^2  - 4}),0\}\ ,
\label{poles}
\end{equation}
\begin{equation}
{\rm zeroes \ \ for \ \ g_{\phi \phi}}:  \ \ \ {{-\chi\pm \sqrt{-3\chi^2+4}}
\over{2\sqrt{3}(\chi^2-1)}} \ .
\label{zeroes}\end{equation}
Thus for $\chi^2>{4/3}$ there are {\it no} zeroes of the metric, but 
there are poles for the values of $\phi$ specified by (\ref{poles}).

Supersymmetric vacua are determined by zeroes of $\partial_\phi W$
(\ref{delw1}). As discussed at the beginning of this section, 
there are at most two, where $\phi$ takes the value:
\begin{equation}
{\rm zeroes\ \ for \ \ \partial_\phi W}: \quad  {{(\chi-\xi)\pm 
\sqrt{\chi^2-\xi\chi+\xi^2-1}} \over
{\sqrt{3}(\xi\chi-1)}} \ .\label{wpzer}
\end{equation}
Note also that the poles  of   $W$, $\partial_{\phi} W$ 
and  $g_{\phi \phi}$
are identical. 

For the parameter range  $\chi^2<{4\over 3}$,   $W$ has
 no poles and thus, due to the relationship 
$\partial_{\phi}^2 W|_{extr} = \textstyle{2\over 3}\, g_{\phi
\phi}\, W|_{extr}$,  one extremum has to be a maximum and
the other one  a minimum. Therefore, the scalar metric $g_{\phi \phi}$ 
is negative
for one value of (\ref{wpzer}) and  corresponds to a non-physical vacuum
($\phi$ becomes tachyonic).

Thus the only physical region for the domain wall solutions
corresponds to $\chi^2>{4\over 3}$. Now the scalar metric
$g_{\phi\phi}$ is always positive definite and $W$ has two real
extrema which are necessarily separated by at least one pole.  Thus
these domain walls interpolate between supersymmetric extrema with the
kink ($\phi$) solution transversing a region where the potential (the
superpotential and the metric) have a pole.  Near the pole the Killing
spinor equations (\ref{killm}) and (\ref{killp}) can be solved
approximately: instead of a typical kink behaviour
$\phi-\phi_{pole}\sim (z-z_{pole})$ (in the case of a finite
potential) now the kink ``slows-down'' and behaves near the pole as
$\phi-\phi_{pole} \sim (z-z_{pole})^{3}$ and the metric coefficient
behaves as $A(z) \sim (z - z_{pole})^{2c}$ where $c$ depends on the
value of $\phi$ at the pole. For $c\ne 0$ the curvature
blows-up~\footnote{We thank R.  Myers for pointing that to
us.}. ($c<0$ for $\xi > {1 \over 2} ( \chi \pm \sqrt{\chi^2 - 4/3} )$
for poles at $\phi= \textstyle{1\over 2}(-\sqrt{3}\chi \pm \sqrt{3
\chi^2 - 4})$.)

The type of the domain wall is specified by the relative signs of $W$
on each side, as well as by the asymptotic behavior of the metric
coefficient.
%For the Randall-Sundrum setup one
%has to approach the Killing horizon asymptotically.  but as we will discuss in
%more detail
%in \cite{270}, the AdS vacua on each side of a BPS domain wall are close
%to the AdS boundary instead the Killing horizon, which was crucial in the
%Rundall-Sundrum setup. 
Typically one encounters Type IV domain wall solutions, i.e. ${\rm
sign}[W_{+\infty}]=-{\rm sign}[W_{-\infty}]$, $A(z)\sim
e^{\sqrt{-{2\over 3}\Lambda_{\pm\infty}}|z|}$ as $|z|\to \infty$, and
at least one pole in-between. (Typical Randall-Sundrum scenario would
correspond to Type II domain walls.)  This is a consequence of the
fact that the supersymmetric extrema are minima of the potential $V$
(\ref{potential}).  Details will be given elsewhere \cite{270}.

For the sake
of concreteness we exhibit a
numerical solution for $\chi =1.4$, $\xi=-0.6$ with the two
supersymmetric minima (\ref{wpzer}) at $ \{-1.0887, -0.1664\}$ sandwiched
between the pole in the middle (poles (\ref{poles}) are at
$\{-1.8980,-0.5269,0\}$). This solution is close to a $Z_2$ symmetric
solution; $W|_{-\infty}= 2.6944$ and $W|_{+\infty}=-2.4953$.  Notice the
``slow-down'' of the kink solution $\phi(z)$ and a power-law ``spike'' of
the metric $A(z)^{-1}$ in the middle of the wall. Moreover, the smoothness of
the scalar is due to the definition as ratio, $X^0$ as well as $X^1$ are
singular at the pole (recall $\phi = X^1/X^0$).

 \begin{figure}
   \vskip0cm
\psfig{figure=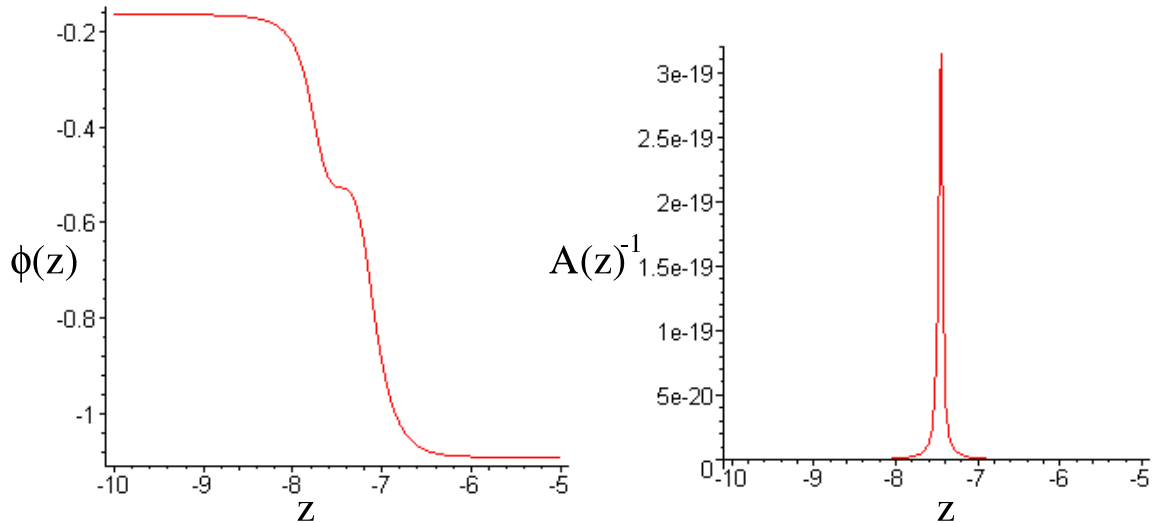,width=6in,angle=0}
   \vskip0.7cm
\caption{The domain wall solution for the scalar field $\phi(z)$ and the metric
coefficient $ A(z)^{-1}$ is depicted for the choice of parameters $\chi=1.4$
and
$\xi=-0.6$. Note the ``slow-down'' of the kink  and  a ``spike'' of the metric
coefficient $A(z)^{-1}$ in the region in the middle of the wall.}
  \end{figure}

%%%%%%%%%%%%%%%%%%%%%%%%%%%%%%%%%%%%%%%%%%%%%%%%%%%%

\section{Conclusions}

%%%%%%%%%%%%%%%%%%%%%%%%%%%%%%%%%%%%%%%%%%%%%%%%%%%

The specific realization of supersymmetric domain walls with asymptotically AdS
space-times, in the simplest
five-dimensional supergravity, demonstrates a number of interesting
features.  The superpotential $W$ as a function of a single scalar can
have at most two extrema, but there is no smooth  flow possible while
demanding that the scalar metric remains positive ($g_{\phi\phi} >0$).
Two AdS vacua with positive scalar metric have to be separated by a pole
in the superpotential and the corresponding domain wall represents a
supergravity  solution that interpolates between the two branches.
Despite this singularity, a stable kink solution exists with the scalar
field ``slowing-down'' mildly in the region crossing the pole. 

Within a more general  framework of five-dimensional N=2 $U(1)$ 
gauged supergravity 
we derived  the Killing spinor equations for static 
domain wall solutions with asymptotic AdS space-times. For  regular
solutions in this class  the  wall tension and the cosmological constants of
the supersymmetric AdS vacua are related.  Such a relationship  is  therefore 
a consequence of the  supersymmetry and
not an artifact of fine-tuning.
 As another by-product we see that the
domain wall world is flat and  supersymmetric, i.e. along with the
massless graviton there is an accompanying gravitino.   The
hypermultiplets of D=5 gauged supergravity could potentially play a
role of matter on the domain wall, a subject of further
investigations. The break-down of supersymmetry
(by either of the vacua) would ensure that the non-extreme walls would
become expanding bubbles (see the analysis given in for
non-supersymmetric domain walls in D=4 in \cite{200} and for a
somewhat related analysis in D=5 in \cite{212,170,171}.)

\bigskip

{\bf Acknowledgments}

We would like to thank S.~Gubser and  R.~Myers  for helpful comments, P.~
Townsend for correspondence and T. Banks, H. L\" u, A. Tseytlin, S. Gukov
and E. Witten for useful discussions. The work is supported by a DFG
grant, in part by the Department of Energy under grant number
DE-FG03-92-ER 40701 (K.B.), DOE-FG02-95ER40893 (M.C.) and the University
of Pennsylvania Research Foundation (M.C).

%%%%%%%%%%%%%%%%%%%%%%%%%%%%%%%%%%%%%%%%%%%%%%

%
% ---- Bibliography ----
%
% \nocite{*}                   %this uses *everything* in the .bib file
% \bibliography{mirjam99}          %or whatever your .bib file is

\begin{thebibliography}{10}

\bibitem{010}
S.~Ferrara, M.~Porrati, and A.~Zaffaroni, ``N=6 supergravity on {$AdS_5$} and
  the {SU(2,2/3)} superconformal correspondence,'' {\em Lett. Math. Phys.} {\bf
  47} (1999) 255, \href{http://xxx.lanl.gov/abs/hep-th/9810063}{{\tt
  hep-th/9810063}}.

\bibitem{190}
A.~Khavaev, K.~Pilch, and N.~P. Warner, ``New vacua of gauged {N=8}
  supergravity in five-dimensions,''
  \href{http://xxx.lanl.gov/abs/hep-th/9812035}{{\tt hep-th/9812035}}.

\bibitem{180}
A.~Kehagias and K.~Sfetsos, ``On running couplings in gauge theories from type
  {IIB} supergravity,'' {\em Phys. Lett.} {\bf B454} (1999) 270,
  \href{http://xxx.lanl.gov/abs/hep-th/9902125}{{\tt hep-th/9902125}}.

\bibitem{220}
S.~S. Gubser, ``Dilaton driven confinement,''
  \href{http://xxx.lanl.gov/abs/hep-th/9902155}{{\tt hep-th/9902155}}.

\bibitem{020}
D.~Z. Freedman, S.~S. Gubser, K.~Pilch, and N.~P. Warner, ``Renormalization
  group flows from holography supersymmetry and a c-theorem,''
  \href{http://xxx.lanl.gov/abs/hep-th/9904017}{{\tt hep-th/9904017}}.

\bibitem{150}
K.~Behrndt, ``Domain walls of {D=5} supergravity and fixpoints of {N=1} super
  {Yang Mills},'' \href{http://xxx.lanl.gov/abs/hep-th/9907070}{{\tt
  hep-th/9907070}}.

\bibitem{160}
K.~Behrndt, ``{AdS} gravity and field theories at fixpoints,''
  \href{http://xxx.lanl.gov/abs/hep-th/9809015}{{\tt hep-th/9809015}}.

\bibitem{030}
D.~Z. Freedman, S.~S. Gubser, K.~Pilch, and N.~P. Warner, ``Continuous
  distributions of {D3}-branes and gauged supergravity,''
  \href{http://xxx.lanl.gov/abs/hep-th/9906194}{{\tt hep-th/9906194}}.

\bibitem{230}
A.~Brandhuber and K.~Sfetsos, ``Nonstandard compactifications with mass gaps
  and {N}ewton's law,'' \href{http://xxx.lanl.gov/abs/hep-th/9908116}{{\tt
  hep-th/9908116}}.

% \bibitem{040}
% L.~Randall and R.~Sundrum, ``Out of this world supersymmetry breaking,''
%   \href{http://xxx.lanl.gov/abs/hep-th/9810155}{{\tt hep-th/9810155}}.

\bibitem{050}
L.~Randall and R.~Sundrum, ``A large mass hierarchy from a small extra
  dimension,'' \href{http://xxx.lanl.gov/abs/hep-ph/9905221}{{\tt
  hep-ph/9905221}}.

\bibitem{070}
L.~Randall and R.~Sundrum, ``An alternative to compactification,''
  \href{http://xxx.lanl.gov/abs/hep-th/9906064}{{\tt hep-th/9906064}}.

\bibitem{060}
N.~Arkani-Hamed, S.~Dimopoulos, G.~Dvali, and N.~Kaloper, ``Infinitely large
  new dimensions,'' \href{http://xxx.lanl.gov/abs/hep-th/9907209}{{\tt
  hep-th/9907209}}.

\bibitem{240}
J.~Lykken and L.~Randall, ``The shape of gravity,''
  \href{http://xxx.lanl.gov/abs/hep-th/9908076}{{\tt hep-th/9908076}}.

\bibitem{250}
C.~Csaki and Y.~Shirman, ``Brane junctions in the {Randall-Sundrum} scenario,''
  \href{http://xxx.lanl.gov/abs/hep-th/9908186}{{\tt hep-th/9908186}}.

\bibitem{260}
A.~E. Nelson, ``A new angle on intersecting branes in infinite extra
  dimensions,'' \href{http://xxx.lanl.gov/abs/hep-th/9909001}{{\tt
  hep-th/9909001}}.

\bibitem{211}
A.~Lukas, B.A.~Ovrut, K.S.~Stelle and D.~Waldram,
``The universe as a domain wall,''
Phys.\ Rev.\ {\bf D59} (1999) 086001 hep-th/9803235.

\bibitem{210}
A.~Lukas, B.~A. Ovrut, K.~S. Stelle, and D.~Waldram, ``Heterotic {M} theory in
  five-dimensions,'' {\em Nucl. Phys.} {\bf B552} (1999) 246,
  \href{http://xxx.lanl.gov/abs/hep-th/9806051}{{\tt hep-th/9806051}}.

\bibitem{212}  
A.~Lukas, B.A.~Ovrut and D.~Waldram,
``Boundary inflation,'' hep-th/9902071.


\bibitem{261}
L.~Girardello, M.~Petrini, M.~Porrati and A.~Zaffaroni,
``The Supergravity Dual of N=1 Super Yang-Mills Theory,''
\href{http://xxx.lanl.gov/abs/hep-th/99090047}{{\tt hep-th/9909047}}.


\bibitem{080}
M.~Gunaydin, G.~Sierra, and P.~K. Townsend, ``The {N=2} {M}axwell-{E}instein
  supergravity theories: Their compact and noncompact gaugings and {J}ordan
  algebras,''. To appear in Proc. of Workshop on Supersymmetry and its
  Applications, Cambridge, England, Jun 24 - Jul 12, 1985.

\bibitem{081}
M.~Gunaydin, G.~Sierra, and P.~K. Townsend, ``More on d = 5
  {M}axwell-{E}instein supergravity: Symmetric spaces and kinks,'' {\em Class.
  Quant. Grav.} {\bf 3} (1986) 763.


\bibitem{120}
M.~Cveti\v c, S.~Griffies, and S.-J. Rey, ``Static domain walls in {N=1}
  supergravity,'' {\em Nucl. Phys.} {\bf B381} (1992) 301--328,
  \href{http://xxx.lanl.gov/abs/hep-th/9201007}{{\tt hep-th/9201007}}.

\bibitem{130}
M.~Cveti\v c and H.~H. Soleng, ``Supergravity domain walls,'' {\em Phys. Rept.}
  {\bf 282} (1997) 159, \href{http://xxx.lanl.gov/abs/hep-th/9604090}{{\tt
  hep-th/9604090}}.


\bibitem{270}
K.~Behrndt, M.~Cveti\v c and S.S.~Gubser, work in progress.

\bibitem{140}
M.~Cveti\v c and S.~Griffies, ``Domain walls in {N=1} supergravity,''
  \href{http://xxx.lanl.gov/abs/hep-th/9209117}{{\tt hep-th/9209117}}.

\bibitem{280}
M.~Cveti\v c, R.~Davis, S.~Griffies, and H.~H. Soleng, ``Cauchy horizons,
  thermodynamics and closed timelike curves in planar supersymmetric
  space-times,'' {\em Phys. Rev. Lett.} {\bf 70} (1993) 1191--1194,
  \href{http://xxx.lanl.gov/abs/hep-th/9210123}{{\tt hep-th/9210123}}.

\bibitem{200}
M.~Cveti\v c, S.~Griffies, and H.~H. Soleng, ``Nonextreme and 
ultraextreme domain
  walls and their global space-times,'' {\em Phys. Rev. Lett.} {\bf 71} (1993)
  670--673, \href{http://xxx.lanl.gov/abs/hep-th/9212020}{{\tt
  hep-th/9212020}}.

\bibitem{100}
S.~Coleman and F.~De Luccia, ``Gravitational effects on and of vacuum decay,''
  {\em Phys. Rev.} {\bf D21} (1980) 3305.

\bibitem{170}
N.~Kaloper, ``Bent domain walls as brane-worlds,''
  \href{http://xxx.lanl.gov/abs/hep-th/9905210}{{\tt hep-th/9905210}}.

\bibitem{171}
H.B.~Kim and H.D.~Kim,
``Inflation and Gauge Hierarchy in Randall-Sundrum Compactification,''
hep-th/9909053.


\end{thebibliography}
% \bibliographystyle{utphys}   %if you use utphys.bst

\providecommand{\href}[2]{#2}\begingroup\raggedright\endgroup

%%%%%%%%%%%%%%%%%%%%%%%%%%%%%%%%%%%%%%%%%%%%%%%%%%%%%%%%

\end{document}